%% file: main_CLEAN.tex
\documentclass[twocolumn,prb,notitlepage,floats,superscriptaddress,amsmath,amssymb]{revtex4-2}

\usepackage[version=3]{mhchem} 
\usepackage{bm}
\usepackage[utf8]{inputenc}
\usepackage[T1]{fontenc}
\usepackage{graphicx}
\usepackage{upgreek}
\usepackage{color}
\usepackage{ulem}
\usepackage{hyperref} 
\hypersetup{colorlinks,citecolor=blue, filecolor=blue ,linkcolor=blue , urlcolor=blue, pdftex}
\usepackage{sidecap}
\usepackage{amssymb}
\usepackage{braket}
\usepackage[caption=false]{subfig}
\usepackage{soul}

\def \FUW{Institute of Experimental Physics, Faculty of Physics, University of Warsaw, 02-093 Warsaw, Poland}
\def \Praga{Department of Condensed Matter Physics, Faculty of Mathematics and Physics, Charles University, CZ-121 16 Prague, Czech Republic}
\begin{document}

\title{Exciton spectrum in atomically thin monolayers: \\
The role of hBN encapsulation}

\author{Artur O. Slobodeniuk}
\email{aslobodeniuk@karlov.mff.cuni.cz}
\affiliation{\Praga}
\author{Maciej R. Molas}
\email{maciej.molas@fuw.edu.pl}
\affiliation{\FUW}

\begin{abstract}
The high-quality structures containing semiconducting transition metal dichalcogenides (S-TMDs) monolayer (MLs) required for optical and electrical studies are achieved by their encapsulation in hexagonal BN (hBN) flakes. 
To examine the effect of hBN thickness in these systems, we consider a model with an S-TMD ML placed between a semi-infinite in the out-of-plane direction substrate and complex top cover layers: a layer of finite thickness, adjacent to the ML, and a semi-infinite in the out-of-plane direction top part. 
We obtain the expression for the Coulomb potential for such a structure. 
Using this result, we demonstrate that the energies of excitonic $s$ states in the structure with WSe$_2$ ML change significantly for the top hBN with thickness less than 30 layers for different substrate cases, such as hBN and SiO$_2$.
For the larger thickness of the top hBN flake, the binding energies of the excitons are saturated to their values of the bulk hBN limit. 
\end{abstract}

\maketitle
\section{Introduction \label{sec:Intro}}
The properties of excitons, electron-hole ($e$-$h$) pairs bounded by Coulomb force, in two-dimensional (2D) monolayers (MLs) of semiconducting transition metal dichalcogenides (S-TMDs) are remarkably modified due to a significant change in the
Coulomb interaction between charge carriers in such 2D crystals~\cite{Tawinan2012, Ashwin2012, Diana2013}.
The excitons are characterized by the energy spectrum, composed in analogy to the hydrogen series as of the ground (1$s$) 
and excited (2$s$, $2p$, 3$s$ $\dots$) states.
Although excitonic states of the $s$-type are observable in the linear optical spectra of S-TMD MLs, $i.e.$, photoluminescence~\cite{Liu2019spectrum, Chen2019, Molas2019Energy, Kapuscinski2021, Sell2022}, transmission~\cite{Stier2018, Goryca2019, AroraTrion}, and reflectance contrast~\cite{Chernikov2014, Molas2019Energy, Gerber2019}, the excitonic states of the $p$- and $d$-types can be seen in non-linear experiments performed on S-TMD MLs, $i.e.$, second harmonic generation or two-photon absorption~\cite{Ye2014, He2014, Wang2015, Kusaba2021}.
It turns out that the energy spectrum of $s$-type states in these atomically-thin semiconductors does not reproduce the conventional Rydberg series of a 2D hydrogen atom~\cite{Macdonald1986, Koteles1988}. 
The main reason for that is the dielectric inhomogeneity of the \mbox{S-TMD} structures, $i.e.$, MLs surrounded by dielectric materials. 
While the Coulomb interaction scales as $\propto 1/\varepsilon r$ with the dielectric response of the surrounding medium $\varepsilon$ at large $e$-$h$ distances $r$, it appears to be significantly weakened at short $e$-$h$ distances due to exceptionally strong dielectric screening within the ML plane.
Consequently, the energy spectrum of excitons in S-TMD MLs and hence their binding energy, defined as the energy difference between the electronic band gap and the ground 1$s$ state, can be strongly modified by the used surrounding media of different dielectric responses.

The influence of the surrounding dielectric on the excitonic ladder has been studied both experimentally and theoretically~\cite{Chernikov2014, Stier2016, Raja2017, Stier2018, Molas2019Energy, Goryca2019, Hsu2019, Jensen2020, Bieniek2022, Shi2022, Truong2022, AroraTrion}.
Note that the theoretical approaches rely mostly on the {\it ab initio}~\cite{Gerber2018, Latini2015, Rosner2016, Florian2018}, 
as well as the combination of the {\it ab initio} and analytical methods~\cite{Andersen2015,VanTuan2018}.
In the latter case, the results of {\it ab initio} simulations have been used as input parameters for the analytical models, 
which are called quantum electrostatic heterostructure (QEH) models. 
Such the incoming parameters are dielectric functions of the multilayer van der Waals heterostructure~\cite{Andersen2015,Latini2015}, momentum dependent matrix elements of the screened Coulomb interaction and the band structure of the valence and conduction bands~\cite{Florian2018}, or even the modified Coulomb potential $V_{3\chi}(\rho)$ in Ref.~\cite{VanTuan2018}. 
However, in all these cases the calculation of the excitons' energies requires large computational powers. 
Therefore, the QEH model, which takes into account all the basic characteristics of the heterostructure, but requires less computational resources to calculate the excitonic spectrum, is still needed.

Note that the current approach to obtain the highest-quality S-TMD MLs is based on their encapsulation in flakes of atomically flat hexagonal BN (hBN).
It results, in particular, in a substantial narrowing of excitonic resonances approaching the homogeneous linewidth limit~\cite{Ajayi2017, Cadiz2017, Wierzbowski2017}, that allows to identify precisely their spectrum.   
Consequently, it is of the utmost importance to perform theoretical calculations of the thickness influence of the surrounding media on the excitons spectra in S-TMD MLs within the aforementioned QEH model.

In this work, we investigate theoretically the energy spectrum of free excitons in S-TMD MLs encapsulated in between a semi-infinite in the out-of-plane direction bottom substrate and complex top cover layers consisting of two parts: a layer of finite thickness $L$, adjacent to the ML, and semi-infinite in the out-of-plane direction top part with the aid of generalization of the Rytova-Keldysh potential.
We demonstrate that the energies of the excitonic $s$ states in such a system with the WSe$_2$ ML are strongly modified when the thickness of the top hBN layers decreases below about 30 layers. 
In addition, it results in a significant reduction in excitonic binding energy ($E_\mathrm{b}$) of almost 40\% in the transition from the sample without the top hBN layer ($E_\mathrm{b}$=256~meV) to the one with an infinite thickness of the top hBN layer ($E_\mathrm{b}$=165~meV). 
The similar behavior of the binding energies as a function of the thickness of the top hBN layer has been observed for the other type of substrates.

The paper is organized as follows. 
In Sec.~\ref{sec:Coulomb}, we present the theoretical framework for the calculation of the effective Coulomb potential in a non-homogeneous planar system, presented in Fig.~\ref{fig:scheme}. 
We analyze the analytical expression for the potential in momentum as well as in coordinate space, as a function of the parameters of the system. 
In Sec.~\ref{sec:potential_hbn} we consider the particular case of the hBN substrate and hBN top flake of finite thickness $L$, and calculate the corresponding effective Coulomb potential for this case. 
Using the obtained potential, we calculate in Sec.~\ref{sec:spectrum_hbn} the energy ladder of the excitons for the case of the WSe$_2$ monolayer as a function of the number of layers of the top hBN flake. 
In Sec.~\ref{sec:summary}, we summarize all findings. 
Moreover, the Supplementary Material (SM) presents additional calculations that take into account the discrete structure of the top hBN layer.
Using this result, we obtain the spectrum of the excitons in WSe$_2$ monolayer with mono- and bilayer top hBN layer and compare the result found within the model proposed in the main text. 
We also study the role of the non-zero distance $\delta$ between the monolayer and the sub- and superstrate on the excitonic spectrum in such a system.

\section{Coulomb potential in the non-homogeneous system: general case}
\label{sec:Coulomb}

\begin{figure}[t]
	\centering
	\includegraphics[width=\linewidth]{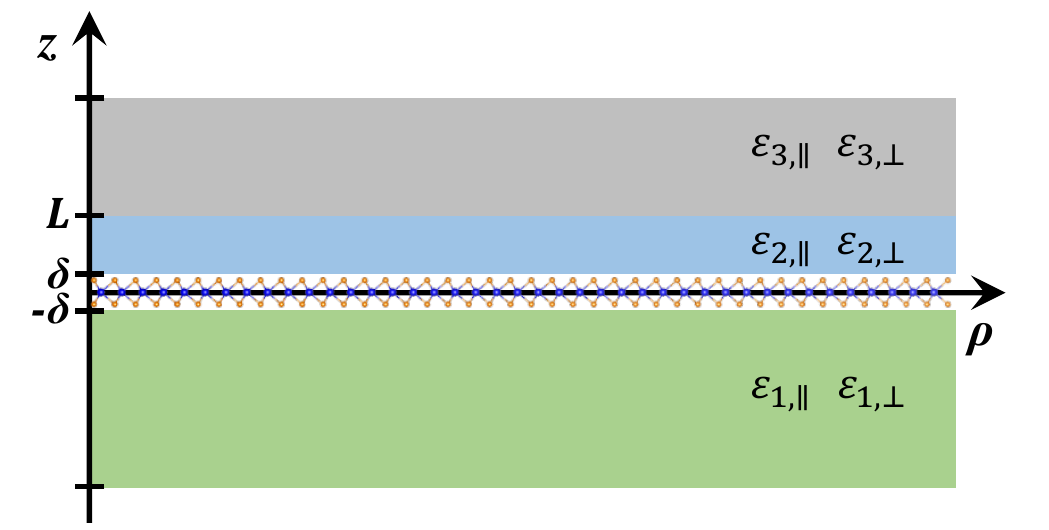}%
	\caption{The schematic illustration of the S-TMD monolayer encapsulated in between a semi-infinite bottom substrate (\mbox{1-st} layer) and a complex top cover layers consisting of two parts: 2-nd layer of the finite thickness, adjacent to the ML and semi-infinite in the out-of-plane direction 3-rd part. } 
	\label{fig:scheme}
\end{figure}

Let us consider the S-TMD ML encapsulated in between a semi-infinite bottom substrate (1-st layer) and complex top cover layers consisting of two parts: 2-nd layer of finite thickness $L$, adjacent to the ML, and semi-infinite in the out-of-plane direction 3-rd part.
A schematic illustration of the studied system is presented in Fig.~\ref{fig:scheme}.
The ML is arranged in the $xy$ plane and is centered in the out-of-plane direction ($z=0$).
The bottom substrate, \mbox{1-st} layer, belongs to the domain $z\in]-\infty, -\delta]$, and is characterized by the in-plane $\varepsilon_{1,\parallel}$ and out-of-plane $\varepsilon_{1,\perp}$ dielectric constants. 
The top (2-nd) layer, next to the ML, unfolds in the range $z\in[\delta, L]$ with the in-plane $\varepsilon_{2,\parallel}$ and out-of-plane $\varepsilon_{2,\perp}$ dielectric constants. 
Finally, the 3-rd top layer spreads over the distance $z\in[L, \infty[$ and is described by the in-plane $\varepsilon_{3,\parallel}$ and out-of-plane $\varepsilon_{3,\perp}$ dielectric constants.    

To find the potential energy between two charges in S-TMD MLs, we solve the following electrostatic problem. 
We investigate the point-like charge $Q$ at the point $\mathbf{r}=(\boldsymbol{\rho},z)=(0,0,0)$ and calculate the electric potential in such a system following Refs.~\cite{Cudazzo2011, Keldysh1979}.
Namely, we analyze four regions: bottom ($z\in]-\infty, -\delta]$), ML ($z\in[-\delta,\delta]$), top finite 
($z\in[\delta, L]$) and overtop ($z\in[L,\infty[$) media, with potentials $\Phi_1(\boldsymbol{\rho},z)$, 
$\Phi(\boldsymbol{\rho},z)$, $\Phi_2(\boldsymbol{\rho},z)$, and $\Phi_3(\boldsymbol{\rho},z)$, respectively. 
These potentials are defined by the Maxwell equations.
It is convenient to present the potentials as a Fourier transform
\begin{align}
\Phi_j(\boldsymbol{\rho},z)=&\frac{1}{(2\pi)^2}\int d^2\mathbf{k} e^{i\mathbf{k}\boldsymbol{\rho}}\Phi_j(\mathbf{k},z), \\
\Phi(\boldsymbol{\rho},z)=&\frac{1}{(2\pi)^2}\int d^2\mathbf{k} e^{i\mathbf{k}\boldsymbol{\rho}}\Phi(\mathbf{k},z).
\end{align}

The $\Phi_j(\boldsymbol{\rho},z)$ potentials for the $j$-th region, where $j=1,2,3$, satisfy Maxwell's equation $\mathrm{div}\,\mathbf{D}_j(\boldsymbol{\rho},z)=0$, which can be written as
\begin{equation}
-\varepsilon_{j,\parallel}\mathbf{k}^2\Phi_j(\mathbf{k},z)+\varepsilon_{j,\perp}
\frac{d^2\Phi_j(\mathbf{k},z)}{dz^2}=0. 
\end{equation}
The solutions of these equations are 
\begin{align}
\Phi_1(\mathbf{k},z)&=B_1e^{\kappa_1 z}  &\text{for~}&~z\in]-\infty, -\delta], \\
\Phi_2(\mathbf{k},z)&=A_2e^{-\kappa_2 z}+B_2e^{\kappa_2 z} &\text{for~}& z\in[\delta,L], \\
\Phi_3(\mathbf{k},z)&=A_3e^{-\kappa_3 z} &\text{for~}& z\in[L,\infty[,
\end{align}
where $\kappa_j=|\mathbf{k}|\sqrt{\varepsilon_{j,\parallel}/\varepsilon_{j,\perp}}=
k\sqrt{\varepsilon_{j,\parallel}/\varepsilon_{j,\perp}}$.

Maxwell's equation in the ML domain, $i.e.$, $z\in[-\delta,\delta]$, reads
$\text{div}\,\mathbf{D}(\boldsymbol{\rho},z)=4\pi Q\delta(\boldsymbol{\rho})\delta(z)$. 
It gives the equation for the potential $\Phi(\mathbf{r},z)$ 
\begin{equation}
\label{eq:coordinate}
\Big[\Delta_\parallel+\frac{d^2}{dz^2}\Big]\Phi(\boldsymbol{\rho},z)=
-4\pi[Q\delta(\boldsymbol{\rho})\delta(z) - \varrho_\text{ind}(\boldsymbol{\rho},z)],
\end{equation}
where $\Delta_\parallel$ is 2D Laplace operator.  
The first term in the right-hand-side of Eq.~(\ref{eq:coordinate}) is the charge density of the charge $Q$, 
localized in the ML plane.  
The second term represents the polarization charge density $\varrho_\text{ind}(\boldsymbol{\rho},z)$, 
induced in the ML by point charge $Q$, which is given by
\begin{equation}
\varrho_\text{ind}(\boldsymbol{\rho},z)=\mathrm{div}\,\mathbf{P}(\boldsymbol{\rho},z).
\end{equation}
Following Ref.~\cite{Cudazzo2011}, we present the polarization in the form
\begin{equation}
\mathbf{P}(\boldsymbol{\rho},z)=\delta(z)\mathbf{P}_\parallel(\boldsymbol{\rho},z=0).
\end{equation}
Using the proportionality between the induced polarization $\mathbf{P}_\parallel(\boldsymbol{\rho},0)$
and the in-plane component of the electric field $\mathbf{E}_\parallel(\boldsymbol{\rho},0)$,  $\mathbf{P}_\parallel(\boldsymbol{\rho},0)=
\chi_\text{TMD}\mathbf{E}_\parallel(\boldsymbol{\rho},0)$, we obtain the expression for the induced charge 
\begin{equation}
\label{eq:indcharge}
\varrho_\text{ind}(\boldsymbol{\rho},z)=-\chi_\text{TMD}\delta(z)\Delta_\parallel\Phi(\boldsymbol{\rho},0).
\end{equation}
Here $\chi_\text{TMD}$ is the 2D polarizability of the S-TMD monolayer~\cite{Cudazzo2011, Berkelbach2013}.
Introducing the screening length parameter $r_0=2\pi\chi_\text{TMD}$, and taking the Fourier transformation of 
Eq.~(\ref{eq:coordinate}) with the induced charge from Eq.~(\ref{eq:indcharge}), one gets
\begin{align}
\label{eq:coordinateFourier}
\Big[\mathbf{k}^2-\frac{d^2}{dz^2}\Big]\Phi(\mathbf{k},z)=4\pi Q\delta(z)-2r_0k^2\delta(z)\Phi(\mathbf{k},0).
\end{align}
This is a linear non-homogeneous differential equation of the second order, which solution can be presented as a sum of the general solution of the homogeneous equation and a particular solution of the non-homogeneous equation (see Ref.~\cite{Kipczak2022} and Supplementary Material) 
\begin{equation}
\label{eq:potential}
\Phi(\mathbf{k},z)=\Psi e^{-k|z|}+Ae^{-k z} + B e^{kz}.
\end{equation}
The $k$-dependent parameters of the potential $\Psi$, $A$, and $B$ are not independent. 
The relations between them are defined from Eq.~(\ref{eq:coordinate}). 
Integrating it over $z$ in the domain $z\in[-\epsilon,\epsilon]$ and then taking the limit $\epsilon\rightarrow 0$, one obtains
\begin{align}
\label{eq:restriction}
[1+r_0k]\Psi+r_0k[A+B]=\frac{2\pi Q}{k}.
\end{align} 

Using the continuity of the potential and $z$ component of the displacement field $\mathbf{D}(\boldsymbol{\rho},z)$ on the boundary
of two adjusted domains, one obtains the set of equations for the parameters $\Psi$, $A$, $B$, $B_1$, $A_2$, $B_2$, $A_3$. 
The boundary conditions for the 1-st and ML domains give relations
\begin{align}
B_1e^{-\kappa_1\delta}&=\Psi e^{-k\delta}+Ae^{k\delta}+Be^{-k\delta},\\
\varepsilon_1 B_1e^{-\kappa_1\delta}&=\Psi e^{-k\delta}-Ae^{k\delta}+Be^{-k\delta}.
\end{align} 
The boundary conditions between the ML and 2-nd domains are described by equations
\begin{align}
A_2 e^{-\kappa_2 \delta}+B_2 e^{\kappa_2 \delta}&=\Psi e^{-k\delta}+Ae^{-k\delta}+Be^{k\delta},\\
\varepsilon_2 A_2 e^{-\kappa_2 \delta}-\varepsilon_2 B_2 e^{\kappa_2 \delta}&=\Psi e^{-k\delta}+Ae^{-k\delta}-Be^{k\delta}.
\end{align}
Finally, the boundary conditions between the 2-nd and 3-rd domains give
\begin{align}
A_2 e^{-\kappa_2 L}+B_2 e^{\kappa_2 L}&=A_3e^{-\kappa_3 L},\\
\varepsilon_2 A_2 e^{-\kappa_2 L}-\varepsilon_2 B_2 e^{\kappa_2 L}&=\varepsilon_3 A_3e^{-\kappa_3 L}.
\end{align}
Here, we introduce $\varepsilon_j=\sqrt{\varepsilon_{j,\perp}\varepsilon_{j,\parallel}}$ for $j=1,2,3$.
Solving these equations together with Eq.~(\ref{eq:restriction}), we obtain the values of the $\Psi$, $A$, and $B$ parameters.
Then substituting them into the expression 
$\Phi(\mathbf{k},z=0)=\Psi+A+B=2\pi Q/k \varepsilon(k)$ we obtain the effective 
in-plane Coulomb potential $\Phi(\mathbf{k},z=0)$. 
Here $\varepsilon(k)$ is the dielectric function of the system 
\begin{widetext}
\begin{align}
\label{eq:dielectric_function}
\varepsilon(k)=kr_0+\frac{1-\Big(\frac{\varepsilon_1-1}{\varepsilon_1+1}\Big)\Big(\frac{\varepsilon_2-1}{\varepsilon_2+1}\Big)
e^{-4 k\delta}-\Big(\frac{\varepsilon_2-\varepsilon_3}{\varepsilon_2+\varepsilon_3}\Big)
\Big[ \Big(\frac{\varepsilon_2-1}{\varepsilon_2+1}\Big)-\Big(\frac{\varepsilon_1-1}{\varepsilon_1+1}\Big)e^{-4k\delta }\Big]e^{-2\kappa_2(L-\delta)}}{\Big[1-\Big(\frac{\varepsilon_1-1}{\varepsilon_1+1}\Big)e^{-2k\delta}\Big]
\Big[1-\Big(\frac{\varepsilon_2-1}{\varepsilon_2+1}\Big)e^{-2k\delta}+\Big(\frac{\varepsilon_2-\varepsilon_3}{\varepsilon_2+\varepsilon_3}\Big)\Big[e^{-2k\delta}-\Big(\frac{\varepsilon_2-1}{\varepsilon_2+1}\Big)\Big]
e^{-2\kappa_2(L-\delta)}\Big]}.
\end{align}
\end{widetext}
The coordinate-dependent potential $\Phi(\boldsymbol{\rho})$ for the considered non-homogeneous system with the dielectric function $\varepsilon(k)$ then reads  
\begin{equation}
\Phi(\boldsymbol{\rho})=Q\int_0^\infty dk\frac{J_0(k\rho)}{\varepsilon(k)}, 
\end{equation}
where $J_0(x)$ is the zeroth Bessel function of the first kind. 
One can see that the information about the studied heterostructure is fully contained in the second part of the expression. 
Note that the expression for the in-plane potential contains only the combinations $\varepsilon_j=\sqrt{\varepsilon_{j,\parallel}\varepsilon_{j,\perp}}$ and $\kappa_2=k\sqrt{\varepsilon_{2,\parallel}/\varepsilon_{2,\perp}}$. 

The expression (\ref{eq:dielectric_function}) simplifies in two important limits. The limit $L\rightarrow\infty$ provides
\begin{equation}
\varepsilon(k)=kr_0-1+\sum_{j=1,2}\frac{1}{1-\Big(\frac{\varepsilon_j-1}{\varepsilon_j+1}\Big)e^{-2k\delta}}.
\end{equation}
This formula interpolates between the case of suspended monolayer $\delta\rightarrow \infty$ ($\varepsilon(k)=kr_0+1$) and
the case of monolayer, encapsulated between two media with dielectric constants $\varepsilon_1,\varepsilon_2$, $\delta\rightarrow 0$
($\varepsilon(k)=kr_0+[\varepsilon_1+\varepsilon_2]/2$). Both limit cases correspond to the so-called Rytova-Keldysh potential in coordinate space, first derived in Refs.~\cite{Keldysh1979,Rytova1967}.

Another limit of Eq.~(\ref{eq:dielectric_function}), which corresponds to the situation with zero distance 
$\delta\rightarrow 0$ between the S-TMD monolayer and the dielectric media, provides
\begin{align}
\label{eq:general_result}
\varepsilon(k)=kr_0+\frac{\varepsilon_1-\varepsilon_2}{2}+\frac{\varepsilon_2}{1+\Big(\frac{\varepsilon_2-\varepsilon_3}{\varepsilon_2+\varepsilon_3}\Big)e^{-2\kappa_2 L}}.
\end{align} 
We consider this dielectric function as the simplest extension of the Rytova-Keldysh model for the case of finite thickness $L$ of the
superstrate.

One can see that the key parameter that regulates the shape of $\varepsilon(k)$ 
is $\tau=\exp(-2\kappa_2L)$.
Namely, the long-wavelength $\lambda(k)=2\pi/k\rightarrow \infty$ and short-wavelength  $\lambda(k)=2\pi/k\rightarrow  0$ limits correspond to $\tau \rightarrow1$ and $\tau\rightarrow 0$ cases, respectively. 
In the long-wavelength limit, the dielectric constant is $\varepsilon(k)\rightarrow (\varepsilon_1+\varepsilon_3)/2$. This result reflects the fact that in this case most of the electric field lines occupy the bottom and second-top regions. In this case, neither the S-TMD monolayer nor the thin first top layer contributes significantly to the dielectric response of the system, due to their small volumes in comparison to the volumes of the other regions. The expression for the potential takes the form
$\Phi(\mathbf{k},z=0)\rightarrow 2\pi Q/(k[\varepsilon_1+\varepsilon_3]/2)$.
It provides the following large distance, $\rho\rightarrow\infty$, the behavior of the potential in the coordinate space $\Phi(\rho,z=0)\rightarrow Q/(\rho[\varepsilon_1+\varepsilon_3]/2)$, which is nothing more than the Coulomb potential of point charge $Q$ placed in between two substrates with dielectric constants $\varepsilon_1$ and $\varepsilon_3$, respectively. 

In the opposite limit, $\lambda(k)\rightarrow 0$, the significant part of the electric lines of the charge occupies the monolayer, the first top layer, and bottom regions. In this case, the effective dielectric constant takes the form $\varepsilon(k)\rightarrow kr_0+(\varepsilon_1+\varepsilon_2)/2$. As one can see, the second top substrate does not give a contribution to the potential. The corresponding limit defines the small distance, 
$\rho\rightarrow 0$, behavior of the
potential $\Phi(\rho,z=0)\rightarrow (\pi Q/2)[\text{H}_0(\rho[\varepsilon_1+\varepsilon_2]/2r_0)-Y_0(\rho[\varepsilon_1+\varepsilon_2]/2r_0)]$.

Finally, note that both coordinate-dependent potentials also correspond to two limits $L\rightarrow 0$ and $L\rightarrow \infty$ of the thickness $L$ of the first top layer. 
Therefore, the potential with a finite value of $L$ interpolates between these two potentials, 
as is depicted in Fig.~\ref{fig:figure1} for particular cases of dielectric constants of the surrounding media.

\section{The Coulomb potential in S-TMD sample: effect of finite thickness of hBN top layer}
\label{sec:potential_hbn}

\begin{figure}[!t]
	\centering
	\includegraphics[width=\linewidth]{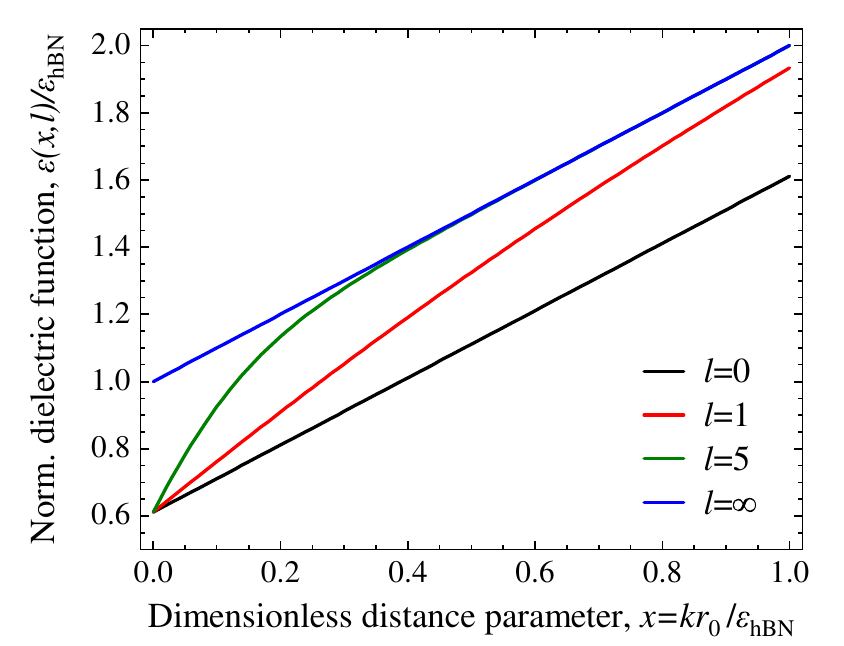}%
	\caption{Normalized dielectric function $\varepsilon(x,l)/\varepsilon_\text{hBN}$ for four values of dimensionless length
        parameter, $l$: $l$=0, $l$=1, $l$=5, and $l$=$\infty$ (black, red, green, and blue curves, respectively) as a function of a dimensionless distance parameter, $x=kr_0/\varepsilon_\text{hBN}$.} 
	\label{fig:dielectric_function}
\end{figure}

We examine the particular case of an S-TMD ML encapsulated in hBN layers, $i.e.$, $\varepsilon_{1,\parallel}=\varepsilon_{2,\parallel}=\varepsilon_{\text{hBN},\parallel}$, $\varepsilon_{1,\perp}=\varepsilon_{2,\perp}=\varepsilon_{\text{hBN},\perp}$, $\varepsilon_{3,\parallel}=\varepsilon_{3,\perp}=1$. 
Following the values available in the literature, we use $\varepsilon_1=\varepsilon_2=\varepsilon_{\text{hBN}}=4.5$, 
and $\kappa_2=k\sqrt{\varepsilon_{\text{hBN},\parallel}/\varepsilon_{\text{hBN},\perp}}\approx 1.098\,k$~\cite{Stier2018}. 
Here we use the high-frequency (infrared) values for the dielectric constants of hBN. This is because the typical frequency scale at which the hBN flake responds to the excitons in the WSe$_2$ monolayer is given approximately by their binding energies of hundreds of meV, see more details in Refs.~\cite{Stier2016,Stier2018,Steinhoff2018}.
Note that $\varepsilon_{3,\parallel}=\varepsilon_{3,\perp}=1$ resemble typical experimental conditions, $i.e.$, the sample is placed in air, vacuum, or gaseous helium. 
Introducing the dimensionless momentum $x=kr_0/\varepsilon_\text{hBN}$ and length $l=\varepsilon_\text{hBN}L/r_0$ parameters, we obtain
the following expression for the effective dielectric constant 
\begin{align}
\varepsilon(x,l)=\varepsilon_\text{hBN}x+\frac{\varepsilon_\text{hBN}}{
1+ \frac{\varepsilon_\text{hBN}-1}{\varepsilon_\text{hBN}+1}
\exp\Big(-2\sqrt{\frac{\varepsilon_{\text{hBN},\parallel}}{\varepsilon_{\text{hBN},\perp}}}xl\Big)}.
\end{align} 
This dielectric function, normalized for $\varepsilon_\text{hBN}$, for different values of the dimensionless thickness $l$ of the top hBN flake is presented in Fig.~\ref{fig:dielectric_function}.

The corresponding effective potential for the considered case $\Phi(\xi)=(Q/r_0)\phi(\xi,l)$, 
as a function of the dimensionless distance $\xi=\rho\varepsilon_\text{hBN}/r_0$, reads
\begin{align}
\label{eq:RKfinal}
\phi(\xi,l)=\int_0^\infty dx\frac{J_0(x\xi)\Big[1+\frac{\varepsilon_\text{hBN}-1}{\varepsilon_\text{hBN}+1}
\exp\Big(-2\sqrt{\frac{\varepsilon_{\text{hBN},\parallel}}{\varepsilon_{\text{hBN},\perp}}}xl\Big)\Big]}{
1+x\Big[1+ \frac{\varepsilon_\text{hBN}-1}{\varepsilon_\text{hBN}+1}
\exp\Big(-2\sqrt{\frac{\varepsilon_{\text{hBN},\parallel}}{\varepsilon_{\text{hBN},\perp}}}xl\Big)\Big]}.
\end{align}
Note that in the limit $l\rightarrow \infty$, $i.e.$, an ML encapsulated in semi-infinite hBN layers, the potential expression is simplified and gives the well-known Rytova-Keldysh potential~\cite{Rytova1967, Keldysh1979}
\begin{align}
\phi(\xi,\infty)=\int_0^\infty dx\frac{J_0(x\xi)}{1+x}=\frac{\pi}{2}\Big[\text{H}_0(\xi)-Y_0(\xi)\Big].
\end{align} 
The other limit $l\rightarrow 0$ corresponds to the case of an ML deposited on a semi-infinite hBN substrate, uncovered from the top.  
The related potential also has the Rytova-Keldysh form  
\begin{align}
\phi(\xi,0)=
\frac{\pi}{2}\Big[\text{H}_0\Big(\frac{\varepsilon_\text{hBN}+1}{2\varepsilon_\text{hBN}}\xi\Big)-
Y_0\Big(\frac{\varepsilon_\text{hBN}+1}{2\varepsilon_\text{hBN}}\xi\Big)\Big].
\end{align} 

\begin{figure}[!t]
	\centering
	\includegraphics[width=\linewidth]{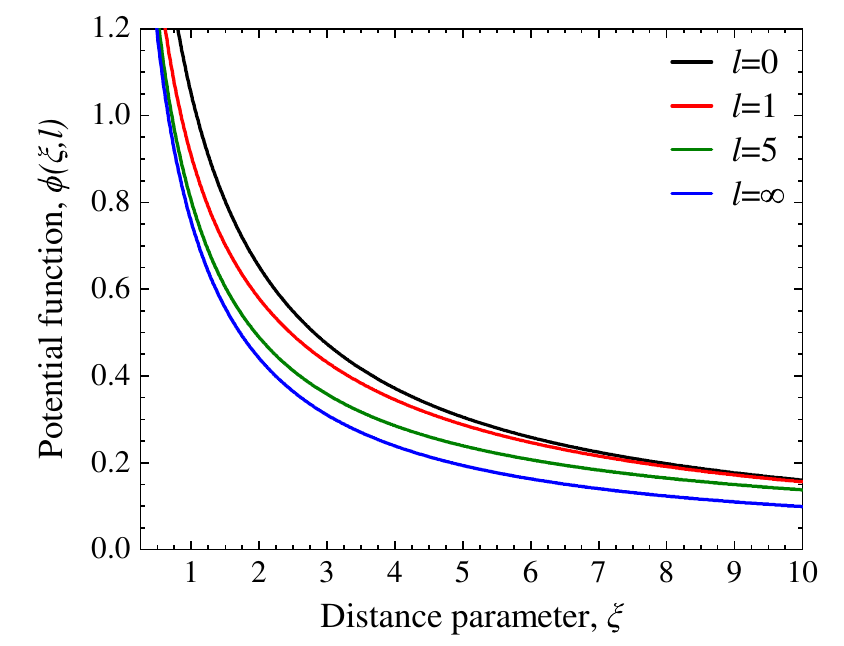}%
	\caption{Potential function $\phi(\xi,l)$ for four values of dimensionless length parameter, $l$: $l$=0, $l$=1, $l$=5, and $l$=$\infty$ as a function of a dimensionless distance parameter, $\xi$.} 
	\label{fig:figure1}
\end{figure}

The evolution of the $\phi(\xi,l)$ potential as a function of a $\xi$ parameter is presented in Fig.~\ref{fig:figure1} for four $l$ values. 
As can be seen in the Figure, the strongest $\phi(\xi,l)$ potential is apparent for $l=0$, while the weakest potential is for the case $l=\infty$, the potential for $l>0$ lies in the region in between of two former potentials. 
The obtained results are in full agreement with the previously reported results~\cite{Raja2017, Stier2018, Truong2022}, where it was shown that an increase in the average dielectric constant of the media surrounding the ML leads to a decrease in the confining potential.

\section{Excitonic spectrum in non-homogeneous system}
\label{sec:spectrum_hbn}

Using the $\phi(\xi,l)$ potential, expressed by Eq.~(\ref{eq:RKfinal}), we can evaluate the energy spectrum of excitons in the investigated structure composed of S-TMD ML as a function of the parameter $l$. 
The corresponding equation for eigenvalues is given by 
\begin{align}
\Big[b^2\frac{1}{\xi}\frac{d}{d\xi}\Big(\xi\frac{d}{d\xi}\Big)+2b\phi(\xi,l)+\epsilon\Big]\psi(\xi)=0,
\end{align}    
where we introduced $b=\hbar^2\varepsilon_\text{hBN}^2/(\mu e^2r_0)$ and $E=Ry^*\epsilon$.
$Ry^*=\mu e^4/(2\hbar^2\varepsilon_\text{hBN}^2)$ is an effective Rydberg energy and $\psi(\xi)$ represents the wave function of an exciton.
$\mu=m_em_h/(m_e+m_h)$ is the reduced mass of the exciton ($e$-$h$ pair) with the effective electron ($m_e$) and hole ($m_h$) masses and $e$ represents electron's charge. 

\begin{figure}[!t]
	\centering
	\includegraphics[width=1\linewidth]{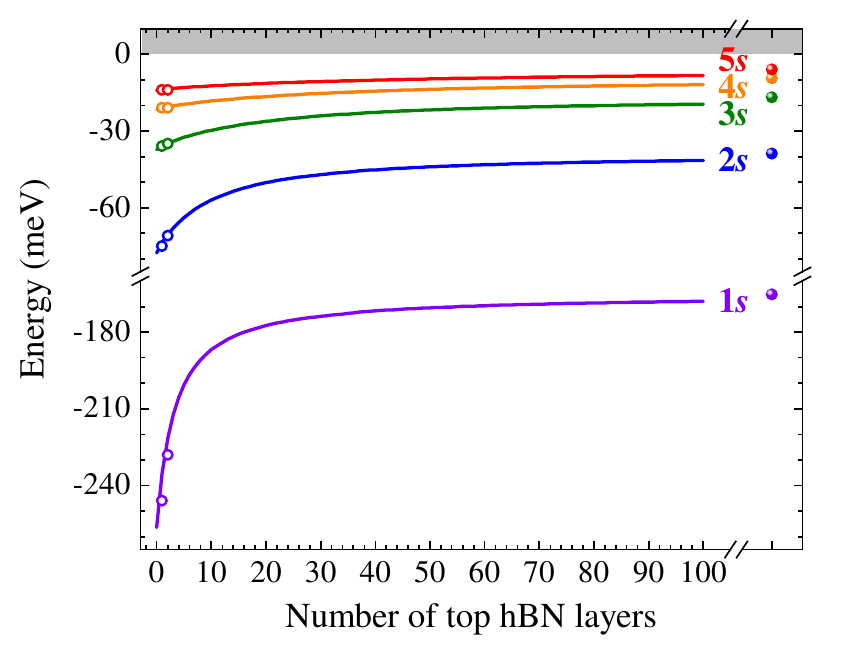}%
	\caption{Energy spectrum of $s$ excitonic state in the WSe$_2$ ML encapsulated in hBN layers as a function of the thickness 
                of the top hBN layer (colored online). The full points correspond to the case of a semi-infinite thickness of top hBN.
                The colored open points represent the energy spectrum of excitons calculated individually for the mono- and bilayer top hBN (see SM for details). Note that the thickness of the bottom hBN layer is semi-infinite in both cases.
                The gray-shaded region represents the infinity of states above the bandgap energy.} 
	\label{fig:spectrum}
\end{figure}

Let us examine the case of WSe$_2$ ML with \mbox{$r_0=4.5$\,nm~\cite{Berkelbach2013}} and $\mu=0.21\,m_0$~\cite{Molas2019Energy}, where $m_0$ is electron's mass. 
It gives $b\approx 1.13357$ and $Ry^*\approx 141$\,meV.
For the WSe$_2$ ML $r_0=4.5$\,nm, the dimensionless parameter $l$ corresponds to the $l$\,nm of the thickness of the top flake of hBN. 
Taking into account that the distance between the layers in hBN (in other words, the thickness of hBN ML) is $d=0.33$\,nm~\cite{Bin2015}, we conclude that $l=1$ corresponds to 3 layers of hBN. 

Note that in the case of the few-layer top hBN flake, when its discreteness can play an important role,
the phenomenological continuous model of the hBN medium considered here should be additionally verified.
To do it, we perform separately the numerical analysis for the ML and BL of the top hBN in the Supplementary Material (SM).

The calculated energy spectra of an exciton for the ground (1$s$) and four excited (2$s$ -- 5$s$) states as a function of the thickness of the top hBN layer for the aforementioned continuous model as well as the discrete one for the thinnest layers are presented in Fig.~\ref{fig:spectrum}.
Due to the observed evolutions in the Figure, three main points can be raised:
(i) the continuous model provides a very good method for calculating the exciton spectrum, even in the case of the extremely thin top hBN flake of about 1-2 layers; 
The largest discrepancy is observed between the homogeneous model and the ML of top hBN for the energy of 1$s$ state of about 5$\%$.  
(ii) the energies of excitonic states are subjected to the most significant variations for the thinnest top hBN layers with thicknesses below about 30 layers. 
For thicker top hBN layers, the corresponding excitonic energies are almost fixed; 
(iii) the thickness effect of the top hBN layer is the largest for the ground 1$s$ state of the exciton with its substantial reduction when the number of excitonic states is increased.
The maximum change of the 1$s$ energy of about 91~meV between two limits: without the top hBN layer and with its infinite thickness. 
The analogous differences of the 2$s$, 3$s$, 4$s$, and 5$s$ states are of the order of 39~meV, 20~meV, 13~meV, and 8~meV, respectively.
\begin{center}
\begin{table}[!t]
\centering
\caption{Calculated binding energies of excitons ($E_\mathrm{b}$) in WSe$_2$ ML encapsulated in hBN layers for selected numbers of the top hBN layer.}
\label{tab:tab1}
\begin{tabular}{ccccccccc}
 \hline\hline 
 Number of \\ top hBN layers & 0 & 3 & 6 & 10 & 20 & 40 & 100 & $\infty$  \\ 
 \hline 
  $E_\mathrm{b}$ (meV) & 256 & 212 & 197 & 187 & 177 & 172 & 168 & 165 \\ 
 \hline  
 \end{tabular}
\end{table}
\end{center}
We can focus on analyzing the excitonic binding energy ($E_\mathrm{b}$, defined as the energy difference between the electronic bang gap and the ground 1$s$ state.
The dependence of the $E_\mathrm{b}$ energy in the WSe$_2$ ML encapsulated in the hBN layers for selected numbers of the top hBN layer is summarized in Table~\ref{tab:tab1}.
The experimentally measured binding energies of excitons in WSe$_2$ MLs encapsulated in hBN flakes is of about 170~meV~\cite{Stier2018, Chen2019, Molas2019Energy, Kapuscinski2021}.
Besides the structures with the hBN encapsulation investigated experimentally differ from the one analyzed in this work, $e.g.$, a WSe$_2$ ML sandwiched between 10 nm thick hBN deposited on the core of a single-mode optical fiber~\cite{Stier2018}, the theoretically calculated binding energies are in very good agreement with the experimental ones.
Using our approach, the excitonic binding energy can be changed by almost 40\% in the transition from the sample without the top hBN layer ($E_\mathrm{b}$=256~meV) to the one with the infinite thickness of the top hBN layer ($E_\mathrm{b}$=165~meV). 
This reveals that the thickness of the surrounding media of S-TMD MLs also plays a crucial role in the modification of the exciton energy spectrum in S-TMD MLs in addition to the engineering of the surrounding dielectric environment, $i.e.$, the encapsulation of an ML in media characterized by dielectric constants~\cite{Raja2017}.

To fulfill our results we consider two additional cases which can be realized in the experiment: {\it i)} the case of the SiO$_2$ substrate, $i.e.$, $\varepsilon_1=2.1$; {\it ii)} the case of the suspended S-TMD monolayer together with the thick hBN flake, $i.e.$, $\varepsilon_1=1$. 
We repeat the calculation of the exciton spectrum in the S-TMD monolayer for both cases and compare them with the previous results in the SM. 
Finally, we study the general case, presented by Eq.~(\ref{eq:dielectric_function}), with a non-zero distance gap $\delta$ between the S-TMD monolayer and the sub- and superstrate, see SM for details.   

\section{Summary \label{sec:summary}}
We obtained the generalization of the Rytova-Keldysh potential for the heterostructures which consists of the substrate, the monolayer S-TMD, the top hBN flake of finite thickness, and the overtop superstrate. 
Using the analytical interpretation of the potential, we calculated the energy spectrum of the excitonic states in the S-TMD monolayer placed on a semi-infinite hBN substrate and covered with a top hBN layer with thickness $L$. 
We presented that the binding energies of the excitons can be significantly modified due to screening effects and as a function of thickness $L$.  
For WSe$_2$ ML in such a structure, we demonstrated that the energies of the excitonic states are substantially adjusted for the thinnest top hBN layers with thicknesses below about 30 layers. 
For the thickness of the top layer larger than 30 layers, the binding energies of the excitons saturate to the excitons energies in the bulk hBN case.

Additionally, we have found that the thickness effect of the top hBN layer is the largest for the ground 1$s$ state of the exciton. 
It results in a significant reduction in excitonic binding energy that can be changed by almost 40\% in the transition from the sample without the top hBN layer to the one with an infinite thickness of the top hBN layer.
The proposed model may be applicable to other 2D layered materials in which the screening effects on the excitonic spectrum play an essential role, $e.g.$, 2D perovskites~\cite{Baranowski2020}.

\section{Acknowledgments \label{Acknowledgements}}
The work has been supported by the National Science Centre, Poland (grant no. 2018/31/B/ST3/02111) and 
by the Czech Science Foundation (project GA23-06369S).

\bibliographystyle{apsrev4-2}
\bibliography{biblio}

\newpage
\onecolumngrid
\setcounter{figure}{0}
\setcounter{section}{0}
\renewcommand{\thefigure}{S\arabic{figure}}
\renewcommand{\thesection}{S\Roman{section}}
\include{SI_clean}

\end{document}

%% file: SI_CLEAN.tex
	\begin{center}
	{\large Supplementary Material for "Exciton spectrum in atomically thin monolayers: \\ 	The role of hBN encapsulation"}
\end{center}

\section{The modified Rytova-Keldysh potential for the case of ultrathin top hBN flake}

In our study, we modeled the top layer of hBN with its thickness $L$ and the macroscopic in- ($\varepsilon_{\text{hBN},\parallel}$) and out-of-plane ($\varepsilon_{\text{hBN},\perp}$) dielectric constants of hBN, $i.e.$, in the limit of the continuous hBN medium.  
However, in the case of the ultrathin top hBN flake, where it consists of only a few layers of hBN monolayers, the use of such a limit is doubtful. 
In order to find the limits of applicability of the continuous model, we derive the modified Rytova-Keldysh potential with a few layers of the hBN. 

To do it, we consider the modification of the system studied in the main text. 
The bottom semi-infinite substrate remains unchanged.
Thus, it is characterized by in- ($\varepsilon_{1,\parallel}$) and out-of-plane ($\varepsilon_{1,\perp}$) dielectric constants and occupies the domain $z\in]-\infty, -\delta]$. 
The S-TMD monolayer is placed in the plane $z=0$ with its in-plane polarizability $\chi_\text{TMD}$. 
The first top hBN flake consists of $N$ layer and has a thickness $L=Nd$. 
Here $d=0.33$\, nm is the distance between the hBN layers in a bulk hBN crystal. 
The $j$th layer of the hBN flake is placed in the $z_j=jd$ plane. 
Each hBN layer is characterized by the in-plane polarizability $\chi_\text{hBN}$. 
The second top layer with the in- ($\varepsilon_{3,\parallel}$) and out-of-plane ($\varepsilon_{3,\perp}$) dielectric constants occupies the domain $z\in[L+\delta, \infty[$. 

In order to find the potential energy between two charges in the S-TMD monolayer, we solve the following electrostatic problem.
We consider the point-like charge $Q$ at the point $(0,0,z')$, where $0<z'\neq z_j <L$ and calculate the potential in such a system following~\cite{Cudazzo2011}.
Similarly to the previous case, we consider 3 regions: bottom semi-infinite medium with the potential
$\Phi_1(\boldsymbol{\rho},z)$, second top semi-infinite layer with the potential $\Phi_3(\boldsymbol{\rho},z)$, 
and the space between them with the potential $\Phi_2(\boldsymbol{\rho},z)$, where we introduced the in-plane 
vector $\boldsymbol{\rho}=(x,y)$.
Using the cylindrical symmetry of the problem, we present the potentials in the form
\begin{equation}
\label{eq:fourier}
\Phi_j(\boldsymbol{\rho},z)=\frac{1}{(2\pi)^2}\int d^2\mathbf{k} e^{i\mathbf{k}\boldsymbol{\rho}}\Phi_j(\mathbf{k},z)
\end{equation}

The Maxwell's equation $\text{div}\,\mathbf{D}_{1,3}=0$ for 1-st and 3-rd regions ($j=1,3$) reads
\begin{equation}
-\varepsilon_{j,\parallel}\mathbf{k}^2\Phi_{j}(\mathbf{k},z)+\varepsilon_{j,\perp}\frac{d^2\Phi_j(\mathbf{k},z)}{dz^2}=0.
\end{equation}
The solutions to these equations are
\begin{align}
\Phi_1(\mathbf{k},z)&=Ae^{\kappa_1 z},  &\text{for}& \, z\in]-\infty,-\delta] , \\
\Phi_3(\mathbf{k},z)&=Be^{-\kappa_3 z}, &\text{for}& \, z\in [L+\delta, \infty[,
\end{align}
where $\kappa_j=|\mathbf{k}|\sqrt{\varepsilon_{j,\parallel}/\varepsilon_{j,\perp}}=
k\sqrt{\varepsilon_{j,\parallel}/\varepsilon_{j,\perp}}$.

The equation for the potential $\Phi_2(\mathbf{r},z)$ takes the form
\begin{align}
\label{eq:second_potential}
\Delta_\parallel\Phi_2(\boldsymbol{\rho},z)+&\epsilon_\perp\frac{d^2\Phi_2(\boldsymbol{\rho},z)}{dz^2}= \nonumber \\ &-4\pi[Q\delta(\boldsymbol{\rho})\delta(z-z') - \varrho_\text{ind}(\boldsymbol{\rho},z)],
\end{align}
where $\Delta_\parallel$ is 2D Laplace operator. 
We introduced the out-of-plane dielectric constant ($\epsilon_\perp$) of the hBN flake with $N>1$. 
In the case $N=1$, one needs to put $\epsilon_\perp=1$.   
The first term on the right-hand side of the equation is the charge density of the charge $Q$.
The second term $\varrho_\text{ind}(\boldsymbol{\rho},z)$ represents the induced charge density
due to the polarization of the hBN and S-TMD layer by charge $Q$.
Following Ref.~\cite{Cudazzo2011}, we present the induced charge in the form
\begin{align}
\varrho_\text{ind}(\boldsymbol{\rho},z)=-&\chi_\text{hBN}\sum_{j=1}^N \delta(z-z_j)\Delta_\parallel\Phi_2(\boldsymbol{\rho},z_j)-
\nonumber \\ -&\chi_\text{TMD}\delta(z)\Delta_\parallel\Phi_2(\boldsymbol{\rho},0).
\end{align}
Using the Fourier transform (\ref{eq:fourier}) of the potential $\Phi_2(\boldsymbol{\rho},z)$ 
we present Eq.~(\ref{eq:second_potential}) as 
\begin{align}
\label{eq:second_potential_k}
\Big[\mathbf{k}^2-\epsilon_\perp\frac{d^2}{dz^2}\Big]\Phi_2(\mathbf{k},z)=&4\pi Q\delta(z-z')-
2r_0k^2\delta(z)\Phi_2(\mathbf{k},0) \nonumber \\
-&2R_0k^2\sum_{j=1}^N\delta(z-z_j)\Phi_2(\mathbf{k},z_j),
\end{align}
where we introduced the in-plane screening lengths $r_0=2\pi\chi_\text{TMD}$ and $R_0=2\pi\chi_\text{hBN}$ for 
S-TMD and hBN monolayers, respectively.
Integrating this equation in the regions $z\in[-\epsilon, \epsilon]$, $z\in[z_j-\epsilon, z_j+\epsilon]$, and $z\in[z'-\epsilon,z'+\epsilon]$ one obtains the following conditions in the limit $\epsilon\rightarrow 0$
\begin{align}
\epsilon_\perp\frac{d\Phi_2(\mathbf{k},z)}{dz}\Big|^{+0}_{-0}&=2r_0k^2\Phi_2(\mathbf{k},0), \\
\epsilon_\perp\frac{d\Phi_2(\mathbf{k},z)}{dz}\Big|^{z_j+0}_{z_j-0}&=2R_0k^2\Phi_2(\mathbf{k},z_j),\\
\epsilon_\perp\frac{d\Phi_2(\mathbf{k},z)}{dz}\Big|^{z'+0}_{z'-0}&=-4\pi Q.
\end{align} 
At the points $z\neq 0\,,z_j\,,z'$ the Eq.~(\ref{eq:second_potential_k}) is simplified
\begin{equation}
\Big[\mathbf{k}^2-\epsilon_\perp\frac{d^2}{dz^2}\Big]\Phi_2(\mathbf{k},z)=0
\end{equation}
Therefore, the general solution for $\Phi_2(\mathbf{k},z)$ can be written in the form
\begin{align}
\label{eq:potential2}
\Phi_2(\mathbf{k},z)=&\Psi_0e^{-K|z-z'|}+\sum_{j=1}^N \Psi_je^{-K|z-z_j|} + 
\nonumber \\ +&
\Psi e^{-K|z|}+ \alpha e^{Kz} +\beta e^{-Kz},
\end{align}
where $\Psi_0$,$\Psi_j$,$\Psi$, $\alpha$ and $\beta$ are unknown functions of $K=k/\sqrt{\epsilon_\perp}$.
The aforementioned boundary conditions together with the equation give the following restrictions
\begin{align}
\label{eq:boundary_conditions}
\Psi=&-r_0K\Phi_2(\mathbf{k},0), \nonumber \\
\Psi_j=&-R_0K\Phi_2(\mathbf{k},z_j), \nonumber \\
\Psi_0=&2\pi Q/\epsilon_\perp K.
\end{align}
Considering the boundary conditions for the electrostatic potential and for the out-of-plane component of the displacement field 
 at $z=-\delta$ 
\begin{align}
\Phi_1(\mathbf{k},-\delta)&=\Phi_2(\mathbf{k},-\delta), \\ \frac{\varepsilon_{1,\perp}}{\epsilon_\perp}
\frac{d\Phi_1(\mathbf{k},z)}{dz}\Big|_{z=-\delta}&=\frac{d\Phi_2(\mathbf{k},z)}{dz}\Big|_{z=-\delta},
\end{align}
and at $z=L+\delta$
\begin{align}
\Phi_3(\mathbf{k},L+\delta)&=\Phi_2(\mathbf{k},L+\delta), \\  \frac{\varepsilon_{3,\perp}}{\epsilon_\perp}
\frac{d\Phi_3(\mathbf{k},z)}{dz}\Big|_{z=L+\delta}&=\frac{d\Phi_2(\mathbf{k},z)}{dz}\Big|_{z=L+\delta},
\end{align}
we get the following equations
\begin{widetext}
\begin{align}
&Ae^{-\kappa_1\delta}=\Psi_0e^{-K(z'+\delta)}+\sum_{j=1}^N\Psi_je^{-K(z_j+\delta)}+\Psi e^{-K\delta}+\alpha e^{-K\delta}+\beta e^{K\delta}, \\
&A\frac{\varepsilon_1}{\sqrt{\epsilon_\perp}}e^{-\kappa_1\delta}=\Psi_0e^{-K(z'+\delta)}+\sum_{j=1}^N\Psi_je^{-K(z_j+\delta)}+
\Psi e^{-K\delta}+\alpha e^{-K\delta}-\beta e^{K\delta}, \\
&Be^{-\kappa_3 (L+\delta)}=\Psi_0e^{-K(L+\delta-z')}+\sum_{j=1}^N\Psi_je^{-K(L+\delta-z_j)}+\Psi e^{-K(L+\delta)}+
\alpha e^{K(L+\delta)}+\beta e^{-K(L+\delta)},  \\
&B\frac{\varepsilon_3}{\sqrt{\epsilon_\perp}}e^{-\kappa_3 (L+\delta)}=\Psi_0e^{-K(L+\delta-z')}+\sum_{j=1}^N\Psi_je^{-K(L+\delta-z_j)}+\Psi e^{-K(L+\delta)}-\alpha e^{K(L+\delta)}+\beta e^{-K(L+\delta)}. 
\end{align}
\end{widetext}
where we introduced the short notation $\varepsilon_j=\sqrt{\varepsilon_{j,\perp}\varepsilon_{j,\parallel}}$ for $j=1,3$.
Introducing $a=\varepsilon_1/\sqrt{\epsilon_\perp}$ and $b=\varepsilon_3/\sqrt{\epsilon_\perp}$, and considering the general case $a,b\neq 1$ we remove the parameters $A$ and $B$ from the above equations and obtain
\begin{widetext}
\begin{align}
\Psi_0e^{-K(z'+\delta)}+\sum_{j=1}^N\Psi_je^{-K(z_j+\delta)}+\Psi e^{-K\delta}+\alpha e^{-K\delta}+\Big(\frac{a+1}{a-1}\Big)\beta e^{K\delta}=0,\\
\Psi_0e^{-K(L+\delta-z')}+\sum_{j=1}^N\Psi_je^{-K(L+\delta-z_j)}+\Psi e^{-K(L+\delta)}+\Big(\frac{b+1}{b-1}\Big)\alpha e^{K(L+\delta)}+
\beta e^{-K(L+\delta)}=0. 
\end{align}
\end{widetext}
 
Note that the solution of the aforementioned system of equations for the case $a=1$ and/or $b=1$ can be obtained
from the general solution by taking the corresponding limit. 

We solve the general equation in a few steps. 
First, we write the system of equations in the following form
\begin{align}
\label{eq:alphabeta}
e^{-K\delta}\alpha+\Big(\frac{a+1}{a-1}\Big)e^{K\delta}\beta=&\mathcal{A}, \nonumber  \\
\Big(\frac{b+1}{b-1}\Big)e^{K(L+\delta)}\alpha+e^{-K(L+\delta)}\beta=&\mathcal{B},
\end{align}
where we introduced the following notations $\mathcal{A}=-\Psi_0e^{-K(z'+\delta)}-\sum_{j=1}^N\Psi_je^{-K(z_j+\delta)}-\Psi e^{-K\delta}$, 
$\mathcal{B}=-\Psi_0e^{-K(L+\delta-z')}-\sum_{j=1}^N\Psi_je^{-K(L+\delta-z_j)}-\Psi e^{-K(L+\delta)}$.
Solving the system of equations (\ref{eq:alphabeta}) one gets
\begin{align}
\alpha=\frac{\mathcal{A}e^{-K(L+\delta)}-\mathcal{B}\Big(\frac{a+1}{a-1}\Big)e^{K\delta}}{D}, \\
\beta=\frac{\mathcal{B}e^{-K\delta}-\mathcal{A}\Big(\frac{b+1}{b-1}\Big)e^{K(L+\delta)}}{D}, 
\end{align}
with $D=e^{-K(L+2\delta)}-\Big(\frac{a+1}{a-1}\Big)\Big(\frac{b+1}{b-1}\Big)e^{K(L+2\delta)}$.
Substituting these solutions into (\ref{eq:potential2}), we express $\Phi_2(\mathbf{k},z)$ via $N+2$ parameters $\Psi_0$, $\Psi_j$ and $\Psi$. 
Then, using the latter expression, we calculate the expressions for the potential in $z=0$,  $z=z'$, and $z=z_l,\, l=1,2\dots N$ points
\begin{widetext}
\begin{align}
\label{eq:self_consisted}
\Phi_2(\mathbf{k},0)=&\Psi_0e^{-Kz'}+\sum_{j=1}^N\Psi_je^{-K z_j }+\Psi+\alpha+\beta, \nonumber \\
\Phi_2(\mathbf{k},z')=&\Psi_0+\sum_{j=1}^N \Psi_je^{-K(z_l-z')} + \Psi e^{-K z'}+ \alpha e^{Kz'} +\beta e^{-Kz'},\nonumber  \\
\Phi_2(\mathbf{k},z_l)=&\Psi_0e^{-K(z_l-z')}+\sum_{j=1}^N \Psi_je^{-K|z_l-z_j|} + \Psi e^{-Kz_l}+ \alpha e^{Kz_l} +\beta e^{-Kz_l},
\end{align}
\end{widetext}
which together with (\ref{eq:boundary_conditions}) gives the complete set of equations for the parameters 
$\Psi_0$, $\Psi_j$, and $\Psi$. 
In the system of Eqs.~(\ref{eq:self_consisted}), we considered the particular case $0<z'<z_1<z_2<\dots < z_N=L$ for simplicity. Then we simplify our solutions by taking the following limits $\delta\rightarrow 0$ and $z'\rightarrow 0$, and obtain $\Phi_2(\mathbf{k},z=0)=(2\pi Q/k)v(k)$ which defines the electrostatic potential $\Phi(\rho)$ in the S-TMD plane, which reads
\begin{align}
\Phi(\rho)=Q\int_0^\infty dk J_0(k\rho)v(k). 
\end{align}
We derive the expressions for the case of monolayer ($v_\text{m}$) and bilayer ($v_\text{b}$) hBN flake to demonstrate
the aforementioned algorithm and then estimate the spectrum of excitons for both cases.   

First, we consider the case of monolayer hBN, $i.e.$, $N=1$, $\epsilon_\perp=1$, and hence $a=\varepsilon_1$, $b=\varepsilon_3$ and $K=k$. 
Following the procedure mentioned above, we obtain the result for the monolayer (m) 
\begin{align}
v_\text{m}(k)=
\frac{1-\left(\frac{2kR_0+\varepsilon_3-1}{2kR_0+\varepsilon_3+1}\right)e^{-2dk} }{(kr_0+\frac{\varepsilon_1+1}{2})-
(kr_0+\frac{\varepsilon_1-1}{2})\left(\frac{2kR_0+\varepsilon_3-1}{2kR_0+\varepsilon_3+1}\right)e^{-2dk}}.
\end{align}
Note that this result looks similar to the general result in the main text (see Eq.~23).
The eigenvalue problem for this potential has a form 
\begin{align}
\Big[\widetilde{b}^2\frac{1}{\xi}\Big(\xi\frac{d}{d\xi}\Big)+2\widetilde{b}\int_0^\infty dx J_0(x\xi)v_\text{m}(x)+\epsilon\Big]\psi(\xi)=0.
\end{align}
Here $v_\text{m}(x=kr_0)\equiv v_\text{m}(k)$, $\epsilon=E/Ry$ and $\xi=\rho/r_0$ are the dimensionless energy and coordinate, respectively, with $Ry=\mu e^4/2\hbar^2$ and $\widetilde{b}=\hbar^2/(\mu e^2r_0)$. 
To calculate the spectrum of excitons, we consider the particular case $\varepsilon_1=\varepsilon_\text{hBN}=4.5$,
$\varepsilon_3=1$. 
Using formula (4) from Ref.~\cite{Berkelbach2013} and in-plane dielectric constant of the monolayer hBN 
$\varepsilon_{\text{hBN},\parallel}=4.98$ from \cite{Laturia2018} 
$R_0=(\varepsilon_{\text{hBN},\parallel}-1)d/2$, we obtain $R_0=6.57\,\mbox{\AA}$. 
Note that this screening length is much smaller than, for example, in the WSe$_2$ monolayer $r_0=45\,\mbox{\AA}$ \cite{Berkelbach2013}. 
Solving the eigenvalue equation with $\widetilde{b}\approx 0.056$ and taking into account that $Ry=2.856$\,eV we obtain the following binding energies: $E_1\approx -246$\,meV, $E_2\approx -75$\,meV, $E_3\approx-36$\,meV, $E_4\approx-21$\,meV, $E_5\approx-14$\,meV.
Note that this result is close to the result of the homogeneous model proposed in the main text:  
$E_1\approx -235$\,meV, $E_2\approx -73$\,meV, $E_3\approx-36$\,meV, $E_4\approx-21$\,meV, $E_5\approx-14$\,meV.
The main difference between the two calculated spectra is the binding energy of the $1s$ exciton. 
The smaller binding energy of the $1s$ state in the homogeneous model can be explained by the less effective screening of the Coulomb potential than in the model considered here.    
 
The potential for the case of the bilayer flake $N=2$ $\Phi_2(\mathbf{k},z=0)=(2\pi Q/k\sqrt{\epsilon_\perp})v_\text{b}(k)$ 
with the corresponding out-of-plane dielectric constant $\epsilon_\perp$ is 
\begin{widetext}
\begin{align}
v_\text{b}(k)=\frac{1-\frac{2KR_0(b+2KR_0)}{(KR_0+1)(b+2KR_0+1)}e^{-2dK} +\frac{(KR_0-1)(b+2KR_0-1)}{(KR_0+1) (b+2KR_0+1)}e^{-4dK}}{\frac{a+1}{2}+Kr_0-\frac{[(a+2Kr_0) (b+2KR_0)-1]}{(KR_0+1)(b+2KR_0+1)}KR_0e^{-2dK}+\frac{(KR_0-1) (b+2KR_0-1)}{(KR_0+1)(b+2KR_0+1)}\left(\frac{a-1}{2}+Kr_0\right) e^{-4dK}}.
\end{align}
\end{widetext}
Here $K=k/\sqrt{\epsilon_\perp}$, $a=\varepsilon_1/\sqrt{\epsilon_\perp}$, and $b=\varepsilon_3/\sqrt{\epsilon_\perp}$.
The eigenvalue problem for the case of the bilayer reads 
\begin{align}
\Big[\widetilde{b}^2\epsilon_\perp\frac{1}{\xi}\Big(\xi\frac{d}{d\xi}\Big)+2\widetilde{b}\int_0^\infty dx J_0(x\xi)
v_\text{b}(x)+\epsilon\Big]\psi(\xi)=0.
\end{align}
Here $v_\text{b}(x=kr_0/\sqrt{\epsilon_\perp})\equiv v_\text{b}(k)$, $\epsilon=E/Ry$ and
$\xi=\rho\sqrt{\epsilon_\perp}/r_0$ are the dimensionless energy and coordinate, 
respectively, with $Ry=\mu e^4/2\hbar^2$ and $\widetilde{b}=\hbar^2/(\mu e^2r_0)$. 
For the numerical estimation of the exciton spectrum in this case, we use $\varepsilon_1=4.5$ and $\varepsilon_3=1$ and the result of Ref.~\cite{Laturia2018} for $\epsilon_\perp=2.91$. 
The spectrum of the excitons then reads $E_1\approx -228$\,meV, $E_2\approx -71$\,meV, $E_3\approx-35$\,meV, $E_4\approx-21$\,meV, $E_5\approx-14$\,meV.
It is surprisingly very close to the result of the homogeneous model, described in the main text, 
$E_1\approx-221$\,meV, $E_2\approx-70$\,meV, $E_3\approx-35$\,meV, $E_4\approx -21$\,meV, $E_5\approx -14$\,meV.
Therefore, we conclude that the homogeneous model provides a very good method for calculating the exciton spectrum, even in the case of an extremely thin top hBN flake of about 1-2 layers.

\section{Effective Coulomb potential in the S-TMD flake of finite thickness}
\label{app:vacuum_potential}

We consider a suspended multilayer  crystal of S-TMD as a set of $N$ layers of S-TMD, arranged in parallel to the $xy$ plane.
The position of $j$-th layer is defined by the coordinate $z_j$. We arrange the layers in the following way $0<z_1<z_2<\dots z_N<L$.
As in the previous case, we suppose that a multilayer is polarized in the in-plane direction, with the 2D susceptibility
$\chi_\text{TMD}=r_0/2\pi$. The out-of-plane polarization of the S-TMD crystal is $\epsilon_\perp$.

The electric potential $\Phi(\boldsymbol{\rho},z)$ for the point-like charge $Q$ at the point $\mathbf{r}=(0,0,z')$ of multilayer reads 
\begin{align}
\Big[\Delta_\parallel+\epsilon_\perp\frac{d^2}{dz^2}\Big]\Phi(\boldsymbol{\rho},z)=
&-4\pi Q\delta(\boldsymbol{\rho})\delta(z-z')-\nonumber \\ 
-&2r_0\sum_{j=1}^N \delta(z-z_j)\Delta_\parallel\Phi(\boldsymbol{\rho},z_j).
\end{align}
This equation can be solved with the help of Fourier transform
\begin{align}
\label{Fourier_transform}
\Phi(\boldsymbol{\rho},z)=\frac{1}{(2\pi)^3}\int d^2\mathbf{k}\int_{-\infty}^\infty dq\, e^{i\mathbf{k}\boldsymbol{\rho}+iqz}
\Phi(\mathbf{k},q).
\end{align}
After substitution it into the main equation one gets
\begin{equation}
\label{eq:non_linear_equation}
[k^2+\epsilon_\perp q^2]\Phi(\mathbf{k},q)=4\pi Qe^{-iqz'} -2r_0 k^2
\sum_{j=1}^N e^{-iqz_j}\Phi(\mathbf{k},z_j),
\end{equation}
where 
\begin{equation}
\Phi(\mathbf{k},z_j)=\frac{1}{2\pi}\int_{-\infty}^\infty dq e^{iqz_j}\Phi(\mathbf{k},q).
\end{equation}
Note that a solution of this equation can be found in the $N\rightarrow \infty$ limit, $i.e.$, the case of the infinite (bulk) S-TMD crystal. To do it we rewrite the finite sum in the equation in the form 
\begin{align}
\lim_{N\rightarrow \infty}\sum_{j=1}^N e^{-iqz_j}\Phi(\mathbf{k},z_j)=\frac{1}{d}\Phi(\mathbf{k},q).
\end{align}
Here, we used the parametrization $z_j=A + d (j-1)$ with $d=(B-A)/(N-1)$, with $A/B$ as the coordinate of the bottom/top layer of the multilayer, $d$ is the distance between layers in the crystal. To obtain the aforementioned result we present the sum in the form 
\begin{align}
\sum_{j=1}^N e^{i(q'-q)z_j}=e^{i(q'-q)\frac{(B+A)}{2}}
\frac{\sin\Big[\frac{(q'-q)}{2}dN\Big]}{\sin\Big[\frac{(q'-q)}{2}d\Big]},
\end{align}
then considering $(q'-q)dN/2\sim 1$, $(q'-q)d/2\sim 1/N$,  and using     
\begin{align}
\lim_{N\rightarrow \infty}\frac{\sin\Big[\frac{(q'-q)}{2}dN\Big]}{\Big[\frac{(q'-q)}{2}d\Big]}=
\pi\delta\Big[\frac{(q'-q)}{2}d\Big],
\end{align}
we obtain 
\begin{align}
\lim_{N\rightarrow \infty}\sum_{j=1}^N e^{i(q'-q)z_j}=\pi \delta\Big[\frac{(q'-q)}{2}d\Big].
\end{align}
Then one can rewrite the Eq.~(\ref{eq:non_linear_equation}) in the form 
\begin{equation}
\Big[\Big(1+2\frac{r_0}{d}\Big)k^2+\epsilon_\perp q^2\Big]\Phi(\mathbf{k},q)=4\pi Qe^{-iqz'}.
\end{equation}
This result coincides with the equation for the potential of charge $Q$, placed in the point $z'$ for the crystal
with the in-plane $\varepsilon_\parallel=1+2r_0/d$ and out-of-plane $\varepsilon_\perp=\epsilon_\perp$ 
dielectric constants, see \cite{Berkelbach2013}.  
 
To find the solution of Eq.~(\ref{eq:non_linear_equation}) for the finite number of layers $N$ we divide it on $[\mathbf{k}^2+\epsilon_\perp q^2]$, integrate over $q$ with additional $e^{iqz_m}$ function, evaluate the integral
\begin{equation}
\label{funny_integral}
\int_{-\infty}^\infty dq\frac{e^{iqz}}{k^2+\epsilon_\perp q^2}=\frac{\pi}{k\sqrt{\epsilon_\perp}}e^{-k|z|/\sqrt{\epsilon_\perp}}.
\end{equation}
and obtain the following system of equations for variables $\Phi(\mathbf{k},z_j)$
\begin{equation}
\sum_{j=1}^N \left[\delta_{mj}+\frac{kr_0}{\sqrt{\epsilon_\perp}}e^{-k\frac{|z_m-z_j|}{\sqrt{\epsilon_\perp}}}\right]\Phi(\mathbf{k},z_j)=\frac{2\pi Q}{k\sqrt{\epsilon_\perp}}e^{-k\frac{|z_m-z'|}{\sqrt{\epsilon_\perp}}}.
\end{equation}
This system of the linear equation can be solved analytically for small values of the number of layers $N \sim 1$. 
The solution for the larger numbers $N\gg 1$ can be done numerically. The obtained expressions for the potentials 
in the $z_j$th layer $\Phi(\mathbf{k},z_j)$ are needed to evaluate the effective Coulomb interaction between the electron
and hole localized in the corresponding layers, see Ref.~\cite{Kipczak2022}. 

The alternative way to find the solution of Eq.~(\ref{eq:non_linear_equation}) is to present it in the form of an integral equation
\begin{align}
\Phi(\mathbf{k},q)=\frac{2\pi Q e^{-iqz'}}{k^2+\epsilon_\perp q^2}-
\int_{-\infty}^\infty dq' f(q,q')\Phi(\mathbf{k},q'), 
\end{align}
with the kernel 
\begin{equation}
f(q,q')=\frac{r_0k^2}{k^2+\epsilon_\perp q^2}\frac{1}{\pi}\sum_{j=1}^N e^{-i(q-q')z_j},
\end{equation}
valid for any distribution of the coordinates $z_j$. 

\section{Spectrum of excitons for the different cases of the substrate} 

We evaluate the energy spectrum of excitons in the investigated structure for the case of different substrates as a function of the parameter $l$. 
The eigenvalues equation reads
\begin{align}
\Big[b^2\frac{1}{\xi}\frac{d}{d\xi}\Big(\xi\frac{d}{d\xi}\Big)+2b\phi(\xi,l,\varepsilon_1)+\epsilon\Big]\psi(\xi)=0,
\end{align}    
where we introduced $b=\hbar^2\varepsilon_\text{hBN}^2/(\mu e^2r_0)$ and $E=Ry^*\epsilon$ with 
$Ry^*=\mu e^4/(2\hbar^2\varepsilon_\text{hBN}^2)$, as in the main text. Here $\psi(\xi)$ represents the wave function of an exciton in
terms of dimensionless coordinate $\xi=\rho\varepsilon_\text{hBN}/r_0$.
The dimensionless potential function $\phi(\xi,l,\varepsilon_1)$ has a form 
\begin{widetext}
    \begin{align}
\phi(\xi,l,\varepsilon_1)=\int_0^\infty dx\frac{J_0(x\xi)\Big[1+\frac{\varepsilon_\text{hBN}-1}{\varepsilon_\text{hBN}+1}
\exp\Big(-2\sqrt{\frac{\varepsilon_{\text{hBN},\parallel}}{\varepsilon_{\text{hBN},\perp}}}xl\Big)\Big]}{
x+\frac{\varepsilon_\text{hBN}+\varepsilon_1}{2\varepsilon_\text{hBN}}+\Big[x+ \frac{\varepsilon_1-\varepsilon_\text{hBN}}
{2\varepsilon_\text{hBN}}\Big]\frac{\varepsilon_\text{hBN}-1}{\varepsilon_\text{hBN}+1}
\exp\Big(-2\sqrt{\frac{\varepsilon_{\text{hBN},\parallel}}{\varepsilon_{\text{hBN},\perp}}}xl\Big)}.
\end{align}
\end{widetext}
As one can see for $\varepsilon_1=\varepsilon_\text{hBN}$, it coincides with the previously obtained result, $i.e.$, $\phi(\xi,l,\varepsilon_\text{hBN})=\phi(\xi,l)$, as should be.

\begin{figure}[!t]
	\centering
	\includegraphics[width=1\linewidth]{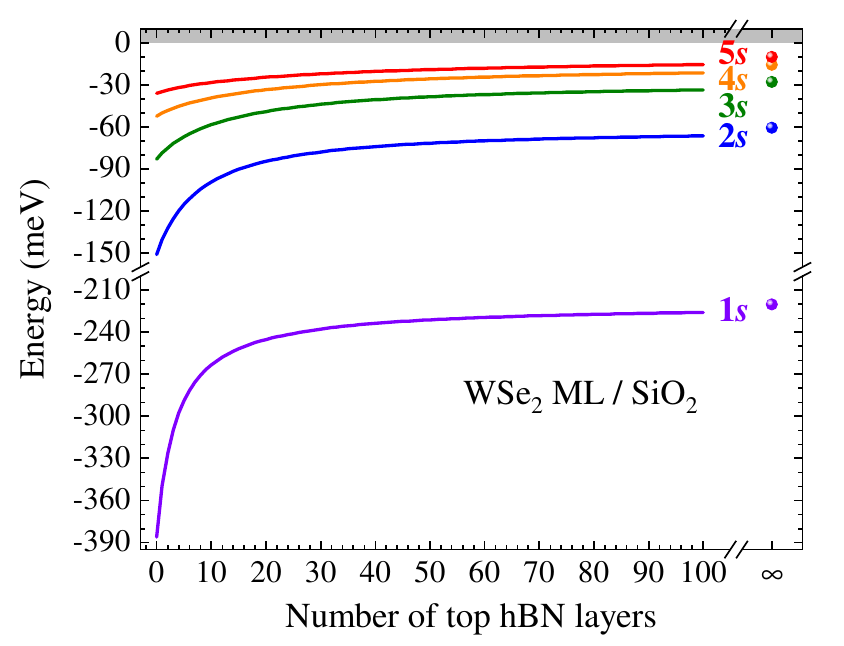}%
	\caption{Energy spectrum of $s$ excitonic state in the WSe$_2$ ML deposited on SiO$_2$ substrate and covered with top hBN layer as a function of its thickness. 
                The gray-shaded region represents the infinity of states above the bandgap energy.} 
	\label{fig:spectrum_sio2}
\end{figure}

\begin{center}
\begin{table}[!t]
\centering
\caption{Calculated binding energies of excitons ($E_\mathrm{b}$) in the WSe$_2$ ML placed on the SiO$_2$ substrate and covered with top hBN layer for selected numbers of the top hBN layer.}
\label{tab:spectrum_sio2}
\begin{tabular}{ccccccccc}
 \hline\hline 
 Number of \\ top hBN layers & 0 & 3 & 6 & 10 & 20 & 40 & 100 & $\infty$  \\ 
 \hline 
  $E_\mathrm{b}$ (meV) & 386 & 310 & 281 & 263 & 245 & 234 & 226 & 220 \\ 
 \hline  
 \end{tabular}
\end{table}
\end{center}

\begin{figure}[!t]
	\centering
	\includegraphics[width=1\linewidth]{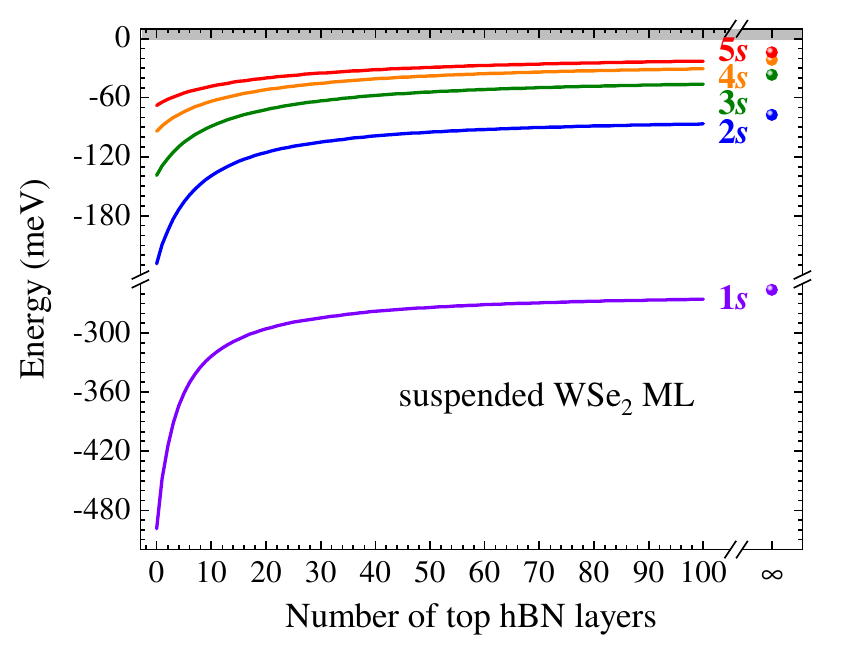}%
	\caption{Energy spectrum of $s$ excitonic state in the suspended WSe$_2$ ML covered with top hBN layer as a function of its thickness. 
                The gray-shaded region represents the infinity of states above the bandgap energy.} 
	\label{fig:spectrum_suspended}
\end{figure}

\begin{center}
\begin{table}[!t]
\centering
\caption{Calculated binding energies of excitons ($E_\mathrm{b}$) in the suspended WSe$_2$ ML covered with top hBN layer for selected numbers of the top hBN layer.}
\label{tab:spectrum_suspended}
\begin{tabular}{ccccccccc}
 \hline\hline 
 Number of \\ top hBN layers & 0 & 3 & 6 & 10 & 20 & 40 & 100 & $\infty$  \\ 
 \hline 
  $E_\mathrm{b}$ (meV) & 499 & 391 & 350 & 324 & 296 & 278 & 266 & 256 \\ 
 \hline  
 \end{tabular}
\end{table}
\end{center}

We solve the eigenvalue equation for the case of SiO$_2$ substrate, $\varepsilon_1=2.1$ \cite{Stier2016} 
and for the case of suspended S-TMD monolayer, $\varepsilon_1=1$.
The calculated energy spectra of an exciton for the ground (1$s$) and four excited (2$s$ $-$ 5$s$) states as a function of the thickness of the top hBN layer for the case of the SiO$_2$ substrate and of the suspended S-TMD monolayer are shown in Figs.~\ref{fig:spectrum_sio2} and \ref{fig:spectrum_suspended}, respectively.
The corresponding dependences of the excitonic binding energy ($E_b$, defined as the energy difference between the electronic bang gap and the ground 1$s$ state) in the WSe$_2$ ML for the case of the SiO$_2$ substrate and of the suspended S-TMD monolayer are summarized Tables~\ref{tab:spectrum_sio2} and \ref{tab:spectrum_suspended}.

\section{Spectrum of excitons for the case of the non-zero distance between monolayer and sub- and superstrate} 

Let us consider the general case of the S-TMD monolayer encapsulated in between different media, presented in Fig.~3 of the main text. 
In this scenario the $k$-depended, and, hence, $\rho$-dependent, effective in-plane Coulomb potential depends on three
distance parameters: the screening length in the plane of the S-TMD monolayer $r_0$, the thickness of the top flake $L$, 
and the distance $\delta$ between the monolayer and the substrate (the same distance is between the monolayer and the top flake). 
As a result, the shape of the Coulomb potential is modified, providing the three-parametric spectrum and wave functions of the
excitons. 
It makes their general analytical consideration quite tricky. 
However, some conclusions can be made 
by studying the limit cases $\delta\rightarrow 0$, $\delta\rightarrow \infty$, $L\rightarrow 0$, etc.

The limit $\delta\rightarrow 0$ is considered in the main text, where it was demonstrated that the spectrum depends on the
dimensionless parameter $L/r_0$. One can conclude that a similar answer can be obtained for small $\delta\ll r_0,L$.
The typical scale of binding energies of the excitons $E_\text{b}$ is suppressed by the dielectric constants of the surrounding media:
$E_\text{b}\propto Ry/(\varepsilon_1+\varepsilon_3)^2$ for $L\rightarrow 0$, and
$E_\text{b}\propto Ry/(\varepsilon_1+\varepsilon_2)^2$ for $L\rightarrow \infty$, see Ref.~\cite{Molas2019} for details.
The opposite limit $\delta\rightarrow\infty$ corresponds to the case of the suspended monolayer, with the dielectric function
$\varepsilon(k)=kr_0+1$. Therefore, the scale of the binding energies of excitons is $E_\text{b}\propto Ry$. 
The intermediate case, where $\delta \sim r_0,L$, interpolates between the two aforementioned limits. Therefore, for fixed parameters
$r_0,L$, the binding energies are increasing with the increasing $\delta$, see also Ref.~\cite{Florian2018}.

\begin{figure}[!t]
	\centering
	\includegraphics[width=1\linewidth]{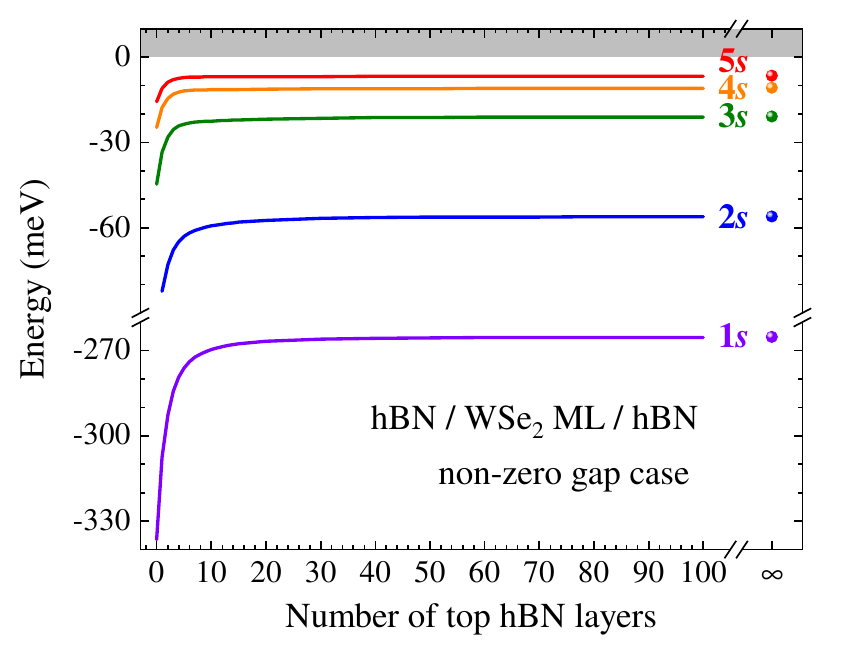}%
	\caption{Energy spectrum of $s$ excitonic state in the hBN-encapsulated WSe$_2$ ML covered with top hBN layer as a function of its thickness for the non-zero distance, $\delta=5\,\mbox{\AA}$, between the ML and sub- and superstrate. 
                The gray-shaded region represents the infinity of states above the bandgap energy.} 
	\label{fig:spectrum_nonzero}
\end{figure}

In order to demonstrate the impact of this parameter on the excitons' spectrum in such a system, we consider the 
WSe$_2$ monolayer, encapsulated in hBN medium, with $\delta=5\,\mbox{\AA}$, as an example.
The corresponding eigenvalue equation has a form 
\begin{align}
\Big[b^2\frac{1}{\xi}\frac{d}{d\xi}\Big(\xi\frac{d}{d\xi}\Big)+2b\phi(\xi,l,\eta)+\epsilon\Big]\psi(\xi)=0,
\end{align}    
where $l=\varepsilon_\text{hBN} L/r_0$, $\eta=\varepsilon_\text{hBN}\delta/r_0$,
$b=\hbar^2\varepsilon_\text{hBN}^2/(\mu e^2r_0)$, $E=Ry^*\epsilon$,
$Ry^*=\mu e^4/(2\hbar^2\varepsilon_\text{hBN}^2)$.
The dimensionless potential function has a form 
\begin{equation}
\phi(\xi,l,\eta)=\int_0^\infty dx J_0(x\xi)/\varepsilon(x,l,\eta),
\end{equation}
with the $\eta$-dependent dielectric function
\begin{widetext}
\begin{align}
\varepsilon(x,l,\eta)=x+
\frac{1-\alpha^2\exp[-4x\eta]-\alpha^2(1-\exp[-4x\eta])
\exp\big[-2\sqrt{\frac{\varepsilon_{\text{hBN},\parallel}}{\varepsilon_{\text{hBN},\perp}}} x(l-\eta)\big]}{\varepsilon_\text{hBN}(1-\alpha \exp[-2x\eta])(1-\alpha \exp[-2x\eta]+\alpha(\exp[-2x\eta]-\alpha)\exp\big[-2\sqrt{\frac{\varepsilon_{\text{hBN},\parallel}}{\varepsilon_{\text{hBN},\perp}}} x(l-\eta)\big])},
\end{align}
\end{widetext}
and $\alpha=(\varepsilon_\text{hBN}-1)/(\varepsilon_\text{hBN}+1)$.
The calculated spectrum, as a function of the number of the layers of the top hBN layer, is presented in Fig.~\ref{fig:spectrum_nonzero}.
The corresponding dependence of the $E_b$ energy is summarized Table~\ref{tab:spectrum_nonzero}.

\begin{center}
\begin{table}[!h]
\centering
\caption{Calculated binding energies of excitons ($E_\mathrm{b}$) in the hBN-encapsulated  WSe$_2$ ML covered with top hBN layer as a function of the number of layers of the top hBN layer for the non-zero distance between the ML and sub- and superstrate, $\delta=5\,\mbox{\AA}$.}
\label{tab:spectrum_nonzero}
\begin{tabular}{ccccccccc}
 \hline\hline 
 Number of \\ top hBN layers & 0 & 3 & 6 & 10 & 20 & 40 & 100 & $\infty$  \\ 
 \hline 
  $E_\mathrm{b}$ (meV) & 336 & 284 & 274 & 270 & 267 & 266 & 265 & 265 \\ 
 \hline  
 \end{tabular}
\end{table}
\end{center}

One can see that the obtained excitonic ladder for the relatively large parameter $\delta$ deviates significantly from the ladder obtained in the main text ($\delta=0$). 
This conclusion is valid also for the general case, presented in Fig.~3 of the main text.
It provides us with the instrument for the analysis of the spectrum in realistic heterostructures. 
Namely, knowing the parameters of the system, one can first calculate the spectrum for the case of $\delta=0$ 
(provided in the main text). If the calculated spectrum significantly deviates from the experimentally observed excitonic ladder, one concludes that there is a non-zero distance gap between the layers, and the more general model, with $\delta>0$, should be considered.